\documentclass[%
 reprint,
 amsmath,amssymb,
prb,
]{revtex4-1}

\usepackage{graphicx}
\usepackage{dcolumn}
\usepackage{bm}
\usepackage{hyperref}
\usepackage{bbold}
\usepackage[caption=false]{subfig}
\usepackage{bm}
\newcommand{\vect}[1]{\boldsymbol{\mathbf{#1}}}
\usepackage{setspace}

\usepackage{float}
\usepackage{paralist}

\begin{document}


\title{Hidden chiral symmetries in BDI multichannel Kitaev chains}

\author{Ant\^{o}nio L. R. Manesco \thanks{A.M. is the corresponding author.}}
 \email[Corresponding author:]{antoniolrm@usp.br}
\affiliation{Lorena Engineering School, University of S\~{a}o Paulo.}
\author{Gabriel Weber}%
 \email{gabrielweber@usp.br}
\affiliation{Lorena Engineering School, University of S\~{a}o Paulo.}%
\author{Durval Rodrigues Jr}
 \email{durval@demar.eel.usp.br}
\affiliation{Lorena Engineering School, University of S\~{a}o Paulo.}





\begin{abstract}

Realistic implementations of the Kitaev chain require, in general, the introduction of extra internal degrees of freedom. In the present work, we discuss the presence of hidden BDI symmetries for free Hamiltonians describing systems with an arbitrary number of internal degrees of freedom. We generalize results of a spinfull Kitaev chain to construct a Hamiltonian with $n$ internal degrees of freedom and obtain the corresponding hidden chiral symmetry. As an explicit application of this generalized result, we exploit by analytical and numerical calculations the case of a spinful 2-band Kitaev chain, which can host up to 4 Majorana bound states. We also observe the appearence of minigap states, when chiral symmetry is broken.

\end{abstract}

\pacs{Valid PACS appear here}
\maketitle


\section{Introduction}

In 1937, Ettore Majorana proposed that a suitable choice for the $\gamma$-matrix representation would lead to real solutions of the Dirac equation, thus implying that the fermions described by these field solutions corresponded to their own antiparticles. \cite{majorana1937teoria} In the past few years, this concept became extremely relevant in the context of Condensed Matter Physics, as Majorana quasiparticle excitations were predicted to emerge in topological superconductors, displaying non-abelian anyonic statistics. This very exotic exchange property has been considered, since then, a very promising route for solving the decoherence problem related to quantum information processing. \cite{kitaev2001unpaired,kitaev2003fault,sato2016topological,read2000paired}

Kitaev, in a seminal paper, introduced a simple toy model, corresponding to a one-dimensional spinless $p$-wave superconductor, capable of hosting Majorana zero energy excitations at both ends.\cite{kitaev2001unpaired} A considerably large number of realistic systems exhibiting such phenomenon were then proposed. The most prominent example consists of a semiconductor nanowire with high spin-orbit coupling in the presence of a magnetic field and in proximity to a $s$-wave superconductor.\cite{lutchyn2010majorana, alicea2010majorana, sau2010generic} Besides the theoretical predictions, there has also been a substantial experimental effort devoted to detecting Majorana bound states in such nanowire heterostructures. \cite{zhang2016ballistic, mourik2012signatures, das2012zero,deng2012anomalous, churchill2013superconductor, finck2013anomalous, albrecht2016exponential, chen2016experimental} In addition, materials with triplet $p$-wave superconductivity, such as organic superconductors and the quasi-one-dimensional K$_{0.9}$Mo$_6$O$_{17}$, \cite{dumitrescu2013topological} as well as other heterostructures such as ferromagnetic nanowires, \cite{dumitrescu2015majorana} were predicted to host Majorana bound states.

Obtaining more realistic realizations of the Physics underlying the Kitaev chain may only be possible with the introduction of internal degrees of freedom, even though it eventually changes the topological classification of the system. For example, for systems such as organic superconductors, quasi-one-dimensional triplet superconductors, like K$_{0.9}$Mo$_6$O$_{17}$, and ferromagnetic nanowires, the relevant internal spin degrees of freedom lead to two different chiral symmetries, one of them characterized by a $\mathbb{Z}$ invariant winding number. \cite{dumitrescu2013topological, dumitrescu2015hidden, dumitrescu2015majorana} On the other hand, in semiconductor nanowire heterostructures, a common feature is the appearance of subbands due to size quantization, requiring the introduction of band mixing terms in the Hamiltonian, which break the DIII chiral symmetry.\cite{lutchyn2011search, lim2012magnetic, lutchyn2011interacting} However, as we have previously shown, hidden chiral symmetries can also be introduced in some limits.\cite{manesco2017onedimensional} Also, multiple Majorana modes counted by a winding number were predicted to appear in long-range hopping systems.\cite{degottardi2013majorana, alecce2017extended} As a matter of fact, correctly accounting for discrete symmetries, such as the chiral symmetry discussed above, is of extreme experimental significance. Particularly, because the breaking of chiral symmetry may lead to the appearance of minigap states, which interfere in the observation of a clear zero-bias peak used as a signature for the presence of Majorana bound states in the system. \cite{zhang2016ballistic, liu2016role} Moreover, theoretical studies of coupled Kitaev chains (Kitaev ladders)\cite{cheng2011majorana, nakai2012time, wakatsuki2014majorana} and multiband systems,\cite{wakatsuki2014majorana, continentino2014topological, potter2011majorana} as well as the recently reported experimental evidence of topological phenomena in a multiband superconductor,\cite{machado2017evidence} corroborate the importance of considering the influence of pairings between internal degrees of freedom on the topological classification of superconductors.

The topological character of a quantum system is uniquely defined by the number of space dimensions and the presence or absence of three discrete symmetries: charge-conjugation, time-reversal and chirality.\cite{altland1997nonstandard,kitaev2009periodic, ryu2010topological} Since in superconductors charge conjugation is manifestly present, the other two, which should occur simultaneously or not at all, are the ones that have to be carefully analyzed when adding internal degrees of freedom. For one-dimensional topological superconductors, several works have suggested the introduction of pseudo-time-reversal operators, \cite{tewari2012topological, dumitrescu2013topological, dumitrescu2015majorana} resulting, for example, in the uncovering of hidden chiral symmetries in spinful systems.\cite{dumitrescu2015hidden} In this work, we propose some conditions to construct Kitaev Hamiltonians with an arbitrary number of internal degrees of freedom and argue that it is also possible to define a hidden BDI chiral symmetry from the superconducting order parameter. These results are applied for a spinful two-band Kitaev chain.

The present paper is organized as follows. In Sec. \ref{classif}, we review the general ideas regarding the classification of one-dimensional topological superconductors, discussing the appropriate topological invariants for a given set of discrete symmetries. In Sec. \ref{spinful-sec}, we briefly review the chiral symmetry leading to the BDI class\cite{dumitrescu2015hidden} and the following geometrical interpretation of the constraints it imposes on the Hamiltonian for the existence of non-trivial topological invariants. In Sec. \ref{n-sec}, we consider, in general, the problem of constructing a Kitaev chain with $n$ degrees of freedom and show how to implement the Nambu representation to find hidden chiral symmetries. In Sec. \ref{2band-sec}, we particularize the previous construction to consider in details the case of a spinful Kitaev chain with two bands. Finally in Sec \ref{conclusion_sec}, we summarize our results and point out some interesting directions and open problems.

\section{Classification of chiral topological superconductors}\label{classif}

Non-trivial topological phases in condensed matter emerge as a consequence of the dimensionality of the system and the discrete symmetries it preserves.\cite{altland1997nonstandard, kitaev2009periodic, ryu2010topological} For superconductors, the mean-field Bogoliubov-de Gennes theory manifestly preserves charge conjugation ($\mathcal{C}$) by construction. Thus, a chiral symmetric system with non-trivial topology necessarily requires time-reversal symmetry, or even a pseudo-time-reversal symmetry, to coexist. \cite{sato2016topological, dumitrescu2015hidden} By pseudo-time-reversal invariance, we mean a symmetry defined by an antiunitary operator that commutes with the Hamiltonian, but does not have the usual physical meaning of a time-reversal. Finally, in a one-dimensional superconductor, given a (pseudo-)time-reversal operator $\mathcal{T}$, the set of possible values for the topological invariant depends on the sign of $\mathcal{T}^2$. \cite{ryu2010topological} In the following we discuss in more details these two cases.

We first consider the case in which $\mathcal{T}_{BDI}^2=1$, corresponding to the BDI class in the ten-fold scheme of classification of topological systems. In this case, the Bloch Hamiltonian can be written in terms of Pauli matrices $\tau_i$ for the particle-hole space as
\begin{equation}
H_k = \vect{h}_k \cdot \vect{\tau}.
\end{equation}
The vector $\vect{h}_k$ defines a topological space ($\mathfrak{T}$) equivalent to a 1-sphere, such that the number of times the vector $\vect{h}_k$ winds around the origin while $k$  goes through the Brillouin zone (BZ) defines distinct topological phases, characterized by a different number of Majorana excitations. In other words, the number of Majorana bound-states can be counted by a topological invariant called the \emph{winding number}, $w \in \pi_1(\mathfrak{T})=\mathbb{Z}$, defined as:\cite{sato2011topology, tewari2012topological}
\begin{equation}
w = \left| \oint_{BZ} \frac{dk}{4\pi i} \text{tr}\left[\mathcal{S}_{BDI}H^{-1}_k\partial_k H_k\right] \right|,
\end{equation}
where $\mathcal{S}_{BDI}$ is the chiral symmetry operator related to the (pseudo-)time-reversal by:
\begin{equation}
\mathcal{S}_{BDI} = i\mathcal{C}\mathcal{T}_{BDI}.
\end{equation} 

On the other hand, systems with a (pseudo-)time-reversal operator that obeys $\mathcal{T}_{DIII}^2=-1$ are in the DIII class. Although we will not make any further comments on how to obtain topological invariants for this class, \footnote{For a more detailed discussion about the topological classes D, BDI and DIII, we refer the interested reader to the works of Budich and Ardonne\cite{budich2013topological} and Sedlmayr \textit{et. al.}\cite{sedlmayr2015majoranas}.} it is important to remark the main difference between systems in the classes BDI and DIII. The presence of a (pseudo-)time-reversal operator that squares to $-1$ implies the presence of Kramer's degeneracy between Majorana excitations. Hence, for one pair of Majoranas to be annihilated, such degeneracy must be broken, requiring that (pseudo-)time-reversal and chiral symmetry be also broken. As a consequence, a DIII system with multiple pairs of Majorana zero modes can have only two distinct topological phases: one with and another without Majoranas. As a result, one must expect a $\mathbb{Z}_2$ invariant instead of $\mathbb{Z}$.

In the following, we focus only on the BDI class, studying how additional internal degrees of freedom may change the behavior of the winding number. To do so, we search for a hidden chiral symmetry, namely an operator $\mathcal{S}$:\cite{sato2016topological, sato2011topology}  
\begin{align}
\mathcal{S} = i \mathcal{C}\mathcal{T} \quad \text{with} \quad \{H_k,\mathcal{S}\}=0, 
\end{align}
defined by the physics of the triplet superconducting order parameter. We start from the idea of hidden chiral symmetry introduced by Dumitrescu \textit{et. al.}\cite{dumitrescu2015hidden} for spinful systems.

\section{The models}
\label{models}

\subsection{A quick review on the spinfull Kitaev chain}
\label{spinful-sec}

We propose a generalized Hamiltonian for a spinfull $p$-wave superconductor considering all possible pairings between the spin channels that are physically compatible with the triplet superconducting state. On Wannier representation, it reads
\begin{align}
\mathcal{H} &= \mathcal{H}_0 + \mathcal{H}_R + \mathcal{H}_{SC}, \label{Wannier_Hamiltonian}\displaybreak[3]\\
\mathcal{H}_0 &= -\sum_{n, \sigma, \sigma'} \mu_{\sigma \sigma'} c_{n\sigma'}^{\dagger} c_{n\sigma} + t_{\sigma \sigma'} c_{n+1\sigma'}^{\dagger} c_{n\sigma} + h.c.,\displaybreak[3]\\
\mathcal{H}_R &= \sum_{n, \sigma, \sigma'} i\lambda_{\sigma \sigma'} c_{n+1\sigma'}^{\dagger} c_{n\sigma} + h.c.,\displaybreak[3]\\
\mathcal{H}_{SC} &= \sum_{n,\sigma,\sigma'} (i\sigma_2 \vect{d} \cdot \vect{\sigma})_{\sigma,\sigma'}c_{n\sigma}^{\dagger}c_{n+1\sigma'}^{\dagger}+h.c.,\displaybreak[3] \label{field-spin-ham}
\end{align}
where $\mu_{\sigma \sigma'}$ and $t_{\sigma \sigma'}$ are the spin dependent chemical potential and the hopping energy, respectively; $i\lambda_{\sigma \sigma'}$ is a purely complex hopping which gives rise to the Rashba spin-orbit coupling and $\vect{d} = (\Delta_1,\Delta_2,\Delta_3)$ is the triplet superconducting order parameter. The fermion field operators $c_{n\sigma}$ and $c_{n\sigma}^{\dagger}$ obey 
\begin{align}
\{c_{n\sigma},c_{m\sigma'}^{\dagger}\} = \delta_{nm}\delta_{\sigma\sigma'}, 
\end{align}
where the indices $n,\ m$ label lattice positions while $\sigma,\ \sigma'$ label the spin projection along the $z$-axis. The set $\{\sigma_{\nu}\}_{\nu=0}^4$ consists of the $2\times 2$ identity matrix and the usual Pauli matrices for the spin space.

For convenience, we rewrite the Hamiltonian \eqref{Wannier_Hamiltonian} in Bloch representation as
\begin{equation}
\mathcal{H} = \int_{BZ} \frac{dk}{2\pi} \; \psi_k^{\dagger}\: H_k \:\psi_k,
\end{equation}
where $BZ$ indicates integration over the Brillouin zone. Using the Nambu representation $\psi_k = (c_k, \mathcal{T}c_k)^T$, $c_k = (c_{k \uparrow}, c_{k \downarrow})^T$, $\mathcal{T} = i\sigma_2 \mathcal{K}$, with $\mathcal{K}$ denoting the complex conjugation operator, we obtain\footnote{From now on, we use Einstein summation convention for repeated indices; greek letters are used for sums starting from 0, while latin letters are reserved for sums starting from 1.}
\begin{align}
H_k &= \tau_3 \otimes (\epsilon_k^{0} \sigma_{0} + \vect{\lambda}_k \cdot \vect{\sigma})+ \tau_0 \otimes (\lambda_k^0 \sigma_0 + \vect{\epsilon}_k \cdot \vect{\sigma}) \nonumber \\
& + \tau_{\phi} \otimes \vect{d}_k\cdot\vect{\sigma},
\label{Hamiltonian}
\end{align}
where $\{\tau_{\nu}\}_{\nu=0}^4$ is the set with the $2\times 2$ identity and the Pauli matrices for particle-hole space; $\tau_{\phi} = \tau_1 \sin \phi + \tau_2 \cos \phi$, $\phi$ is the superconducting phase and 
\begin{align}
[\epsilon_k^{\nu}\sigma_{\nu}]_{\sigma \sigma'} &= -\mu_{\sigma \sigma'} - 2t_{\sigma \sigma'}\cos k,\\
[\lambda_k^{\nu}\sigma_{\nu}]_{\sigma \sigma'} &= 2\lambda_{\sigma \sigma'} \sin k,\\
\vect{d}_k &= \vect{d}\sin k.
\end{align}
We note that
\begin{equation}
-\mu_{\sigma \sigma'} = -\mu \sigma_0 + \vect{B} \cdot \vect{\sigma}
\end{equation}
where $\mu$ is the chemical potential and $\vect{B}$ is a Zeeman field. Also,
\begin{equation}
t_{\sigma \sigma'} = t\sigma_0 + \vect{C} \cdot \vect{\sigma}
\end{equation}
where $t$ is the spin independent hopping energy, $\vect{C}$ is the spin dependent hopping energy.

The Hamiltonian with no spin-dependent hopping was proposed as a realistic model for organic superconductors, such as the quasi-one-dimensional triplet superconductor K$_{0.9}$Mo$_6$O$_{17}$, and ferromagnetic nanowires with zero $s$-wave order parameter. \cite{dumitrescu2013topological, dumitrescu2015hidden, dumitrescu2015majorana} Moreover, it was also pointed out that the parameter choice leads to two possible chiral operators, \textit{i.e.}, unitary operators that anticommute with the Hamiltonian. One is the chiral symmetry related to the DIII classification, $\mathcal{S}_{DIII} = \tau_{\phi + \pi/2} \otimes \sigma_0$, a consequence of the invariance under the physical time-reversal operator defined by $\mathcal{T}_{DIII} = \tau_{0} \otimes i\sigma_2\mathcal{K}$, given $\mathcal{C} = \tau_{\phi+ \pi/2} \otimes \sigma_2\mathcal{K}$. The other is the hidden chiral symmetry associated with the BDI classification, $\mathcal{S}_{BDI} = \tau_{\phi+\pi/2}\otimes \hat{d}\cdot \vect{\sigma}$, $\hat{d} = \vect{d}/|\vect{d}|$, with a corresponding pseudo-time-reversal operator given by $\mathcal{T}_{BDI} = \tau_0 \otimes \left[\hat{d}\cdot \hat{e}_2 + i\left(\hat{d}\wedge\hat{e}_2\right)\cdot \vect{\sigma}\right]\mathcal{K}$.

The conditions for preserving chiral symmetry in a BDI system have an interesting geometric interpretation which we explore next. Imposing chiral symmetry leads to
\begin{equation}\label{spin_chiral_symmetry}
\left\{H_k,\mathcal{S}_{BDI}\right\} = 0 \quad \Rightarrow \quad \left\lbrace
\begin{array}{c}
[\vect{\lambda}_k \cdot \vect{\sigma}, \hat{d}\cdot\vect{\sigma}]=0\\
\lbrace \vect{\epsilon}_k \cdot \vect{\sigma}, \hat{d}\cdot\vect{\sigma} \rbrace=0
\end{array}\right. .
\end{equation}
Since $\left[\epsilon_k^0 \sigma_0, \hat{d}\cdot\vect{\sigma}\right]=0$, the condition \eqref{spin_chiral_symmetry} trivially reduces to:
\begin{align}
\left[\vect{\lambda}_k \cdot \vect{\sigma}, \hat{d}\cdot\vect{\sigma}\right] = 2i\vect{\sigma}\cdot(\vect{\lambda}_k \wedge \hat{d}) = 0 \quad \Rightarrow \quad \vect{\lambda}_k \parallel \hat{d},\label{spinful_parallel_condition} \\
\left\lbrace \vect{\epsilon}_k \cdot \vect{\sigma}, \hat{d}\cdot\vect{\sigma}\right\rbrace=2 \sigma_0 \vect{\epsilon}_k \cdot \hat{d} = 0 \quad \Rightarrow \quad \vect{\epsilon}_k \perp \hat{d}. \label{spinful_perpendicular_condition}
\end{align}
These conditions lock the spin-dependent terms in order to maintain chirality. Finally, it is worth noting that chiral symmetry is only globally realized if $\vect{\epsilon}_k \perp \hat{d}$, $\forall k$, since the $k$-dependency can result in sweet spots for specific values of $k$ due to competition between $\vect{B}$ and $\vect{C}$.

To conclude this section, we remark that, although this construction has been explicitly carried out on the example of spinful systems, a system with any two internal degrees of freedom is described by the same mathematical model, thus, presenting the same ``topology''. Therefore, a spinless system with two bands described in terms of Pauli matrices admits a similar Hamiltonian formulation and invariance under the same hidden symmetry operators, as we demonstrated in a previous work.\cite{manesco2017onedimensional} Based on such arguments, we provide next some general arguments for obtaining hidden BDI chiral symmetries on systems with $n$ internal degrees of freedom and discuss the application of these ideas to a spinfull 2-band Kitaev chain.

\subsection{General construction of a Kitaev chain with $n$ internal degrees of freedom}
\label{n-sec}

The conditions derived in Sec. \ref{spinful-sec} for the  chiral operator originally introduced by  Dumitrescu \textit{et al.} \cite{dumitrescu2015hidden} raises the question of whether it is possible to find similar hidden symmetries for systems with a richer spinorial structure. The idea is to consider a Hamiltonian which is an element of $\mathfrak{su}(2) \times \mathfrak{su}(n)$  (particle hole + other degrees of freedom). It is also necessary to introduce a generalized Nambu representation $\psi_k = (c_k, \mathcal{T}c_k)^T$, where $c_k$ is an element of the spinor representation of $\mathfrak{su}(n)$. Although the construction of $\mathcal{T}$ is highly dependent on the physical meaning attributed to $\mathfrak{su}(n)$ and its representation, some general ideas can be discussed without choosing a specific representation of $\mathcal{T}$. In the next section we will discuss in more details this representation choice for a specific algebra.

Since the Hamiltonian is an element of $\mathfrak{su}(2) \times \mathfrak{su}(n)$, the action of any (pseudo-)time-reversal operator $\mathcal{T} = U_{\mathcal{T}}\mathcal{K}$ ($U_{\mathcal{T}}$ is unitary and $\mathcal{K}$ denotes the complex conjugation) on the generators of $\mathfrak{su}(n)$ divides it in one symplectic subgroup \cite{flint2009sympletic}
\begin{equation}
\mathcal{T}t_a^S\mathcal{T}^{-1} = -t_a^S,
\end{equation}
and one antisymplectic
\begin{equation}
\mathcal{T}t_a^A\mathcal{T}^{-1} = t_a^A.
\end{equation}
Another important point to consider for correctly implementing the Nambu representation is the effect of $\mathcal{T}$ on the $k$-dependency of the Hamiltonian. Thus, we divide the possible terms in symmetric
\begin{equation}
\mathcal{T}\epsilon_k^a\mathcal{T}^{-1} = \epsilon_k^a
\end{equation}
and antisymmetric
\begin{equation}
\mathcal{T}\lambda_k^a\mathcal{T}^{-1} = -\lambda_k^a
\end{equation}
under $\mathcal{T}$.
Taking into account these two effects of the action of $\mathcal{T}$, we propose a general Nambu Hamiltonian
\begin{align}
H_k &= \tau_3 \otimes (\epsilon_k^a t_a^A + \lambda_k^a t_a^S) + \tau_0 \otimes (\epsilon_k^a t_a^S + \lambda_k^a t_a^A) \nonumber \\
&+ \tau_{\phi} \otimes  d_k^a \tilde{t}_a \sin k,
\end{align}
where $\tilde{t}_a$ are the generators of $\mathfrak{su}(n)$ such that $U_{\mathcal{T}}\tilde{t}_a$ are symmetric matrices.\footnote{For example, in the $\mathfrak{su}(2)$ case, $U_{\mathcal{T}} = i\sigma_2$, as explicitly written in \eqref{field-spin-ham}. We note, nonetheless, that changing the spinor representation from the usual to the Nambu removes $U_{\mathcal{T}}$.}

Now we can introduce a hidden chiral symmetry operator similar to the one introduced by Dumitrecu \textit{et al.}\cite{dumitrescu2015hidden} for spinfull systems:
\begin{equation}
\mathcal{S}_{BDI} = \tau_{\phi+\pi/2} \otimes \hat{d}^a \tilde{t}_a
\end{equation}
where $\hat{d}^a$ is the normalized $d^a$ vector such that $\mathcal{S}_{BDI}^2=\mathbb{1}$. Finally, the condition for existence of chiral symmetry, \emph{i.e.}, $\{H_k, \mathcal{S}_{BDI}\}=0$, implies
\begin{align}
[\epsilon_k^a t_a^A + \lambda_k^a t_a^S, \hat{d}^b \tilde{t}_b]=0,\label{cond-n1}\\
\{\epsilon_k^a t_a^S + \lambda_k^a t_a^A, \hat{d}^b \tilde{t}_b\}=0.\label{cond-n2}
\end{align}
These conditions result in a series of constraints on the Hamiltonian, which are analogous to the locking conditions on the spin space obtained in Sec. \ref{spinful-sec}. The chiral operator prohibits some of the coefficients $\epsilon_k^a$ and $\lambda_k^a$ multiplying the generators of $\mathfrak{su}(n)$, \textit{i.e.}, the isospin-dependent terms are locked. However, the geometric interpretation is not completely analogous. The reason lies in the algebraic structure of $\mathfrak{su}(n)$ for an arbitrary $n\geq3$:
\begin{align}
[t_a, t_b] &= if_{ab}^ct_c,\\
\{t_a,t_b\} &= \frac{1}{2n}\delta_{ab} t_0 + g_{ab}^c t_c,
\end{align}
where some of structure constants $f_{ab}^c$ are zero and some $g_{ab}^c$ are non-zero. Thus, the parallel and perpendicular conditions derived in Sec. \ref{spinful-sec} do not hold in general anymore.

Even though we obtained some general conditions for constructing the Hamiltonian and finding hidden chiral symmetries, it is not clear how to apply these results without a specific choice of representation. Thus, we now provide a concrete discussion considering a spinful 2-band system.

\subsection{The spinful 2-band Kitaev chain and its chiral symmetries}
\label{2band-sec}

Following the construction of a Kitaev chain with an arbitrary number of degrees of freedom presented in Sec. \ref{n-sec}, we propose a general Hamiltonian for a spinfull Kitaev chain with two bands. The Hamiltonian is now an element of $\mathfrak{su}(2) \times \mathfrak{su}(4) \cong \mathfrak{su}(2) \times \mathfrak{su}(2) \times \mathfrak{su}(2)$. Denoting the spin (band) subspace by the matrices $\sigma_{\nu}$ ($\rho_{\nu}$), and taking $\mathcal{T} = i\sigma_2 \otimes i\rho_2 \mathcal{K}$, it is straightforward to obtain
\begin{align}
H_k = \tau_3 \otimes &(\epsilon_k^{00} \sigma_0 \otimes \rho_0 + \epsilon_k^{ij} \sigma_i \otimes \rho_j \nonumber \\
&+ \lambda_k^{i0} \sigma_i \otimes \rho_0 + \lambda_k^{0i} \sigma_0 \otimes \rho_i) \nonumber \\
+ \tau_0 \otimes &(\lambda_k^{00} \sigma_0 \otimes \rho_0 + \lambda_k^{ij} \sigma_i \otimes \rho_j \nonumber \\
&+ \epsilon_k^{i0} \sigma_i \otimes \rho_0 + \epsilon_k^{0i} \sigma_0 \otimes \rho_i) \nonumber \\
+ \tau_{\phi} \otimes &d^{ij} \sigma_i \otimes \rho_j \sin k.
\end{align}

Next, we consider the necessary conditions to have the hidden chiral symmetry:
\begin{equation}
\mathcal{S}_{BDI} = \tau_{\phi + \pi/2} \otimes \hat{d}^{ij} \sigma_i \otimes \rho_j,
\end{equation}
where $\hat{d}^{ij}$ is normalized so that $\mathcal{S}_{BDI}^2=\mathbb{1}$. It is evident that $\epsilon_k^{00}$ cannot break chirality, whereas $\lambda_k^{00}$ must be always zero for $\mathcal{S}_{BDI}$ to be preserved, \textit{i.e.}, $\{H_k, \mathcal{S}_{BDI}\}=0$. After collecting the terms with the same matrix structure of the superconducting order parameter, \textit{i.e.}, all terms proportional to $\sigma_i \otimes \rho_j$, the conditions \eqref{cond-n1} and \eqref{cond-n2} lead to
\begin{align}
\epsilon_k^{ij}\hat{d}^{ab}\varepsilon_{ia}^n \varepsilon_{jb}^m &= 0, \label{cond-21}\\
\lambda_k^{ij}\hat{d}^{ij} &= 0. \label{cond-22}
\end{align}
Here, $\varepsilon_{ia}^n$ denotes the totally antisymmetric Levi-Civita tensor in three dimensions. For these terms, the analogy with Sec. \ref{spinful-sec} is direct, because in this case $f_{ab}^c$ are always non-zero and $g_{ab}^c$ are always zero.

\begin{figure}
\subfloat[\label{eigen-1a}]{\includegraphics[width=.4\textwidth]{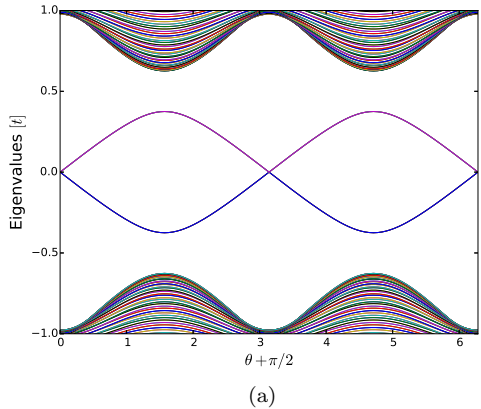}}\\
\subfloat[\label{eigen-1b}]{\includegraphics[width=.4\textwidth]{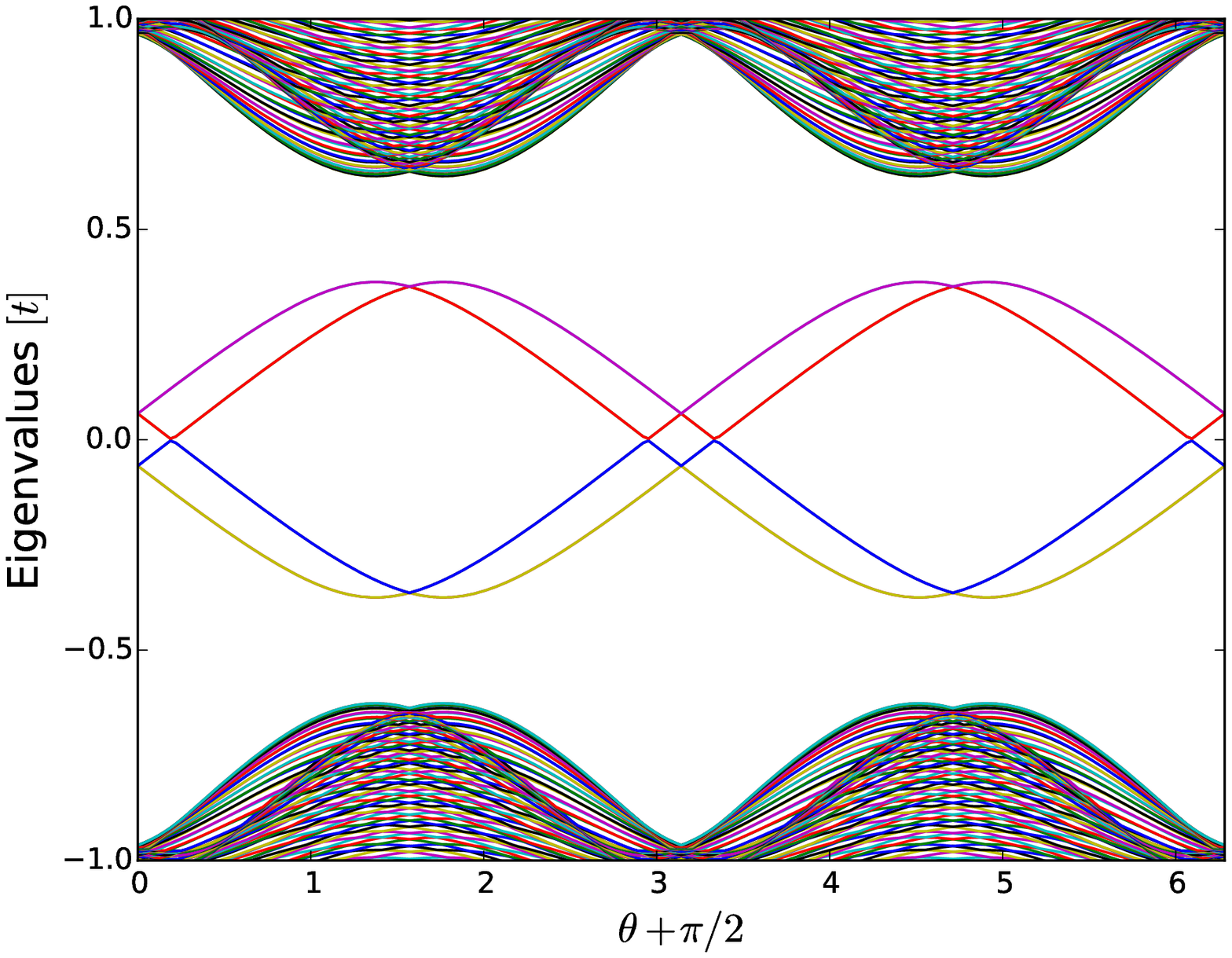}}\\
\caption{Eigenvalues (in units of $t$) for a system described by the Hamiltonian \eqref{ham-num} with 100 lattice positions as a function of $\theta$: (a) $\mu=0$, $\Delta=0.75t$, $m=0.5t$, $\gamma=0$; (b) $\mu=0$, $\Delta=0.75t$, $m=0.5t$, $\gamma=\frac{\pi}{16}$. Non-zero values of $\gamma$ open minigap states for all values of $\theta$, except when breaking chiral symmetry leads to hotspots of zero energy minigap states which are not topologically protected.}
\label{eigen-1}
\end{figure}

To corroborate the results \eqref{cond-21} and \eqref{cond-22} regarding the locking conditions imposed by the superconducting order parameter $\hat{d}^{ab}$, we have performed independent numerical simulations with the \textsc{Kwant} package.\cite{groth2014kwant} For simplicity, we implemented the following representative Hamiltonian:
\begin{equation}
H_k = \tau_3 \otimes (\epsilon_k \sigma_0 \otimes \rho_0 + m \sigma_{\theta} \otimes \rho_{\gamma}) + \tau_{\phi} \otimes \Delta \sigma_1 \otimes \rho_2 \sin k,
\label{ham-num}
\end{equation}
where $\epsilon_k = -\mu -2t\cos k$, $\sigma_{\theta} = \sigma_1 \sin \theta + \sigma_3 \cos \theta$ and $\rho_{\gamma} = \rho_1 \sin \gamma + \rho_2 \cos \gamma$. Chiral symmetry $S_{BDI} = \tau_{\phi+\pi/2} \otimes \sigma_1 \otimes \rho_2$ should be preserved if, and only if, $\sigma_{\theta} = \pm \sigma_1$ and $\rho_{\gamma} = \pm \rho_2$. Therefore, varying the angles $\theta$ and $\gamma$ may lead to the appearance of minigap states when chiral symmetry is broken and of Majorana zero modes when the chirality condition holds. This behavior is explicitly confirmed by Fig. \ref{eigen-1}. For $\gamma=0$ and $\sigma_{\theta} = \pm \sigma_1$, the minigap closes. However, for $\gamma \neq 0$, chiral symmetry is broken for any value of $\theta$ and the minigap only closes when accidental degeneracy emerges. Nonetheless, there is no topological protection in the latter case.

\begin{figure}
\subfloat[\label{phase-1a}]{\includegraphics[width=.3\textwidth]{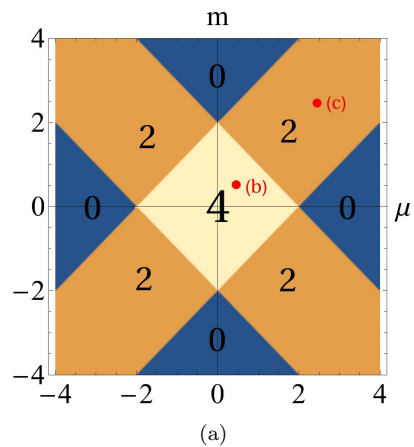}}\\
\subfloat[\label{phase-1b}]{\includegraphics[width=.3\textwidth]{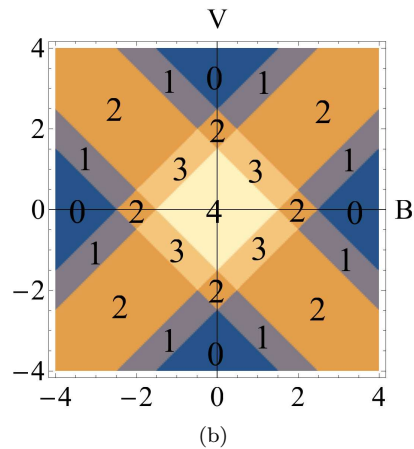}}\\
\subfloat[\label{phase-1c}]{\includegraphics[width=.3\textwidth]{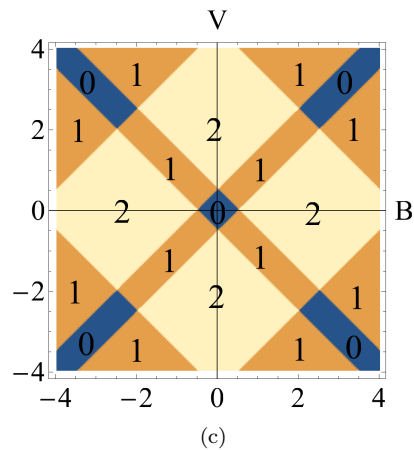}}
\caption{Winding number for a 2-band spinful Kitaev chain. (a) Diagram obtained from Hamiltonian \eqref{ham-num} as a function of $\mu$ and $m$ in units of $t$ with $\Delta = 0.75t$, $\theta = \pi/2$ and $\gamma = 0$. The highlighted points indicate the origin of the diagrams (b) and (c). (b) Diagram obtained from Hamiltonian \eqref{mod-ham-num} with $\Delta = 0.75t$, $\theta = \pi/2$, $\gamma = 0$ and $\mu=0.5$, m$=0.5$. (c) Diagram obtained from Hamiltonian \eqref{mod-ham-num} with $\Delta = 0.75t$, $\theta = \pi/2$, $\gamma = 0$ and $\mu=2.5$, m$=2.5$. It is evident that breaking spin-band symmetry leads to phases with an odd number of Majorana pairs.}
\label{phase-1}
\end{figure}

For the Hamiltonian \eqref{ham-num}, it is also possible to count the number of Majorana zero modes by calculating the winding number. In Fig. \ref{phase-1a}, we show the effect of varying $\mu$ and $m$ on the number of Majorana pairs. As expected, four Majorana pairs are possible. If we increase the absolute values of $\mu$ or $m$, the overlap between these zero modes eventually leads to their annihilation, resulting in lower winding numbers. Finally, we remark that only even winding numbers appear in the phase diagram of Fig. \ref{phase-1a}, which is a feature of a symmetry between spin and band subspaces. This condition will be broken next.

We now consider in more details the influence of the terms proportional to $\sigma_i \otimes \rho_0$ and $\sigma_0 \otimes \rho_i$. One can check that chiral symmetry requires:
\begin{align}
\lambda_k^{i0} \hat{d}^{ab} \varepsilon_{ia}^n = 0,\label{cond-rashba-spin}\\
\lambda_k^{0i} \hat{d}^{ab} \varepsilon_{ib}^m = 0,\label{cond-rashba-band}\\
\epsilon_k^{i0} \hat{d}^{ib} = 0,\label{cond-zeeman-spin}\\
\epsilon_k^{0i} \hat{d}^{ai} = 0. \label{cond-zeeman-band}
\end{align}
It is interesting to note that the previous conditions \eqref{spinful_parallel_condition} and \eqref{spinful_perpendicular_condition} to maintain chiral symmetry on spinfull systems still hold. Namely, \eqref{cond-rashba-spin} implies that $\hat{d}^{ib}$ should be parallel to $\lambda_k^{i0}$ and \eqref{cond-zeeman-spin} means that $\hat{d}^{ib}$ needs to be perpendicular to $\epsilon_k^{i0}$. Also, analogous results \eqref{cond-rashba-band} and \eqref{cond-zeeman-band} hold for band degrees of freedom. Numerical simulations breaking these conditions on the band subspace also resulted on the appearance of minigap states, similar to the ones seen in Fig. \ref{eigen-1}.

To evaluate the effect of breaking spin-band symmetry on the topological phase diagram, we added some of these four terms to the Hamiltonian \eqref{ham-num} according to:
\begin{equation}
H_k \rightarrow H_k + \tau_0 \otimes (B \sigma_3 \otimes \rho_0 + V \sigma_0 \otimes \rho_1). \label{mod-ham-num}
\end{equation}
Here, $B$ denotes a Zeeman field along the $z$-axis and $V$, an analogous contribution to the band subspace, but along the $x$-direction. As expected, odd winding numbers also appear as indicated in Figs. \ref{phase-1b} and \ref{phase-1c}. Hence, the system can indeed host any integer number of Majorana bound states from 0 to 4. Figure \ref{phase-1c} deserves some special care regarding the value of the winding number at the origin. As a matter of fact, in spite of what the diagram may suggest, exactly at the origin, \emph{i.e.}, for $B=V=0$, $w=2$, as consistency with Fig \ref{phase-1a} requires.

Finally, there remains to take into account the effects of $k$-dependent terms, such as Rashba spin-orbit couplings, on the phase diagram. Interestingly, adding such terms to the hamiltonian \eqref{mod-ham-num} don't change the topological phase diagrams in Fig. \ref{phase-1}. Thus, indicating that the Majorana modes are insensitive to them. Nevertheless, for finite systems, the presence of $k$-dependent terms leads to the appearance of minigap states, which became clearer as we shortened the chain. This suggests that away from the continuum limit the very definition of BDI chirality does not hold.

\section{Conclusions}\label{conclusion_sec}

One-dimensional $p$-wave systems with an arbitrary number of internal degrees of freedom allow the emergence of multiple zero energy Majorana excitations at both ends, if a BDI chiral symmetry is preserved. In this paper, we have shown that a hidden chiral symmetry can be derived from the superconducting terms in the Hamiltonian and provided a geometrical interpretation of the constraints imposed on systems that preserve it. This condition locks the isospin-dependent terms of the Hamiltonian by restricting the possible adjoint elements of the $\mathfrak{su}(n)$ representation. We examined in details the consequences of this severe restriction imposed on BDI systems for a spinful 2-band $p$-wave superconductor, in particular, showing that breaking chiral symmetry leads to the emergence of minigap states, that the winding number can assume the values between 0 and 4 and, finally, that odd values of the winding number are only possible when the spin-band symmetry is broken.

Finally, we point out that the construction of pseudo-time-reversal operators for the general case with $n$ degrees of freedom is still a challenging open problem, as well as the appearance of minigap states in finite systems, that the authors wish to revisit in future works.

\section*{Acknowledgments}
The authors would like to thank the Kwant package developers. The work of ALRM was supported by FAPESP grant No. 2016/10167-8. DRJ is a CNPq researcher. The authors wish to thank the financial support by CNPq, FAPESP, and CAPES. The authors would also like to thank Marcelo Hott for useful discussions.

\bibliography{kitaev-2-band}

\end{document}